\documentstyle[aps,twocolumn]{revtex}
\begin{document}
\newcommand{\be}{\begin{equation}}
\newcommand{\ee}{\end{equation}}
\newcommand{\bq}{\begin{eqnarray}}
\newcommand{\eq}{\end{eqnarray}}
\title{Quantum dynamics of two-dimensional vortex pairs with
arbitrary total vorticity}
\author{Vittorio Penna}
\address{ Condensed Matter Section, International Center of
Theoretical Physics, Strada Costiera 11, 34100 -Trieste, Italy
and Dipartimento di Fisica, Politecnico di Torino,
Corso Duca degli Abruzzi 24, I-10129 Torino, Italy.}
\maketitle
\begin{abstract}
Quantum dynamics of a vortex pair is investigated
by considering the pair Hamiltonian within various,
unequivalent algebraic frameworks. First the vortex
pair spectrum is constructed in the standard contest
of the ${\rm e}(2)$-like dynamical symmetry and
its degeneracy is thoroughly examined. 
Then Berry's phase phenomenon is studied through
an ${\rm su}(1,1)$ realization of the pair Hamiltonian when
its parameters are assumed to be time-dependent, whereas 
the Feynman-Onsager quantization conditions are recovered
by means of symmetry arguments within a third approach
based on a magneticlike description of the vortex pair.
Finally, it is shown how recasting the dynamical algebra
in terms of two-particle realizations of both ${\rm su}(2)$
and ${\rm su}(1,1)$ provides the correct approach for
the quantization of the model Hamiltonian accounting
for the pair scattering from a disklike obstacle.
\end{abstract}
%
%
%
\section{Introduction}
An attempt to investigate the quantum dynamics (QD) of
superfluid vortices was performed by Fetter in Ref. \cite{BO}.
He considered a three--dimensional (3D) vortex
model where the vorticity field is nonzero on an array of
parallel strings and assumed that the vortex interactions
were dominated by transversal string oscillations. 
The crucial feature is that the pair of coordinates involved
by the local description of each point of a vortex line are
canonically conjugate allowed to quantize the system via
the canonical procedure.

The complexity of the problems inherent in the quantization
of the vortex field theory together with their extreme
formal character \cite{RARE}
have certainly contributed to discourage, for a long
time, the investigation of the quantum aspects of vortex
dynamics in the context of condensed--matter physics.

The only exception to this statement \cite{AHNS} is represented
by the special case of 2D pointlike massless vortices, which
are derived from the standard field theory of 2D ideal fluids
as the extreme case where the vorticity field is
restricted to a set of isolated points.

Before discussing the motivations that prompt further
investigations of VQD it is interesting to review the work
devoted to 2D vortex dynamics during the last two decades.
The attention raised by its quantum version is principally
due to the vexed question whether point-like vortices might
exhibit, when dynamically quantized, statistics with a fractional
character. A thorough account of this problem can be found in
Refs. \cite{CHM}--\cite{GMS2}
whereas in Ref. \cite {ES} an effective theory of 2D quantum
vortices on a Josephson junctions array is worked out from
the quantum-phase model Hamiltonian.
Further contributions aimed at investigating 2D dynamically
quantized vortices concern the influence of the boundary
effects on VQD \cite{PE1}, the ordering problem arising from
the quantization process \cite{PE2}, the possibility that the
vortex pair inherits anyonlike statistics when its dynamics is
issued from that of two charges in a transverse magnetic field
\cite{LE}, and the emergence of pointlike vortices with quantized
dynamical degrees of freedom from a second quantized
$|\Psi|^4$-field theory \cite{TG}.

Surprisingly, apart from recent developments, such a condensed
survey represents more or less the whole work devoted to the
2D VQD in the past 15 years despite the huge quantity of
theoretical work concentrated, during the same period, on the
2D vortex topic in relation to type-II superconductor
physics \cite{TITI}, the Kosterlitz-Thouless transition theory
\cite{KT1}, \cite{MIN}, and the investigations through Feynman's
variational approach \cite{NAT} on vortex emergence from the
native superfluid background.

A renewed interest for 2D vortex dynamics has been
prompted by the recent experimental developments in the
context of both superfluidity and superconductivity \cite{SUSU}.
The great improvements concerning the measurement
techniques and the observations of microscopic processes
should render quantum aspects of vortex dynamics viable
to experimental detection.

There are at least four experimental situations that are interesting
in respect to VQD. For example, devices where thin films of
superfluid ${\-}^4He$ are adsorbed on porous materials such as Vycor
glasses. The porous structure endows ${\-}^4He$ films with the
multiply-connected geometry of a Riemann surface thus confining
the vortex gas of superfluid films on a network of 2D cylinderlike
connected domains (the surface handles) with very small sizes
\cite{MINA}.
The fact that the special geometry makes vortices interacting
at the mesoscopic scales of the vycor structure is expected
to emphasize the effect of quantizing vortex interactions.
This, in fact, as entailed by the standard form of the
vortex Hamiltonian, depends on the distance between vortex
positions which is the quantity that must be quantized.
The depicted scenery is further complicated by the effect of
the curvature that, as in the case when boundaries confine the
fluid, introduces nonlinear terms in the energy that represent
virtual vortex contributions \cite{PE1}. 

The second situation where vortex structures are important
is the superconductor physics. Point vortices (vortex lines)
arising in 2D (3D) superconductors strongly influence conductance
measurements of the current. The presence in the medium of pinning
centers makes them undergo quantum tunneling phenomena that
strongly affect vortex  movements through the medium \cite{TAN}.
The ensuing liquidlike behavior of the vortex system affects the
supercurrent decay by inducing voltage variations \cite{IEJU}, but
it depends, in turn, on interactions among vortices as well as
on interactions of vortices both with the medium impurities and
with the medium boundaries. Similar effects are observed in
superfluid currents in a ring, while vortex creation phenomena,
possibly due to the quantum tunneling effect, occur when superflows
cross a microscopic orifice \cite{DSSM}.

Finally, vortex dynamics is largely studied in the Josephson
junction arrays \cite{JJA}, where experiments reveal a rich
scenery of phenomena such as high-energetic vortices exhibiting
ballistic motion, vortex-driven voltage turbulence, and vortex
lattice melting.

A characteristic feature of the literature on VQD is the fact
that the attention has been mainly concentrated on the pairs
of identical vortices, namely vortex-vortex (VV) pairs as well
as antivortex-antivortex (AA) pairs. Their deep quantum nature
ensuing from the fact that both VV and AA interactions involve
quantized intervortex distances \cite{CHM} has originated the
idea of characterizing a pair of identical vortices (and, more
in general, a gas of identical vortices) by fractional statistics.
Concerning this point VA pairs were considered uninteresting
since no exotic statistics is expected from distinguishable objects
such as vortices and antivortices. In addition to this, VA pairs
arising in physical systems exhibit a vortex charge which is equal
and opposite to the antivortex charge. This fact implies that the
VA distance is not quantized \cite{PE1} thus making their
dynamics apparently trivial.

VA pairs are instead basic for understanding the
statistical properties of excited superfluids.
These are achieved via mean-field techniques that seem
to entail an intrinsically classic scenery. The standard
renormalization procedure for obtaining scaling laws is illustrative
of this when the habitual assumption is made to consider each VA
pair of the vortex gas as immersed in a dielectric background of VA
pairs with smaller size \cite{MIN}. This implies that both VV and AA
interactions (the very quantum ones) are embodied, and thus averaged,
in the dielectric constant. Since the VA pair Hamiltonian
possesses a sort of natural, semiclassical structure in that it does
not contains any noncommuting quantities, then quantum effects
appear to be a higher--order refinement.

As to the emergence of vortex pairs we wish to recall the main
traits of such a phenomenon within superfluid condensates and
superconductors.
The neutrality characterizing both of them at low temperature favors
the occurrence of VA pairs when temperature is raised. These, in fact,
violate the neutrality just locally and are energetically favored.
On the contrary individual VV and AA pairs represent disfavored
excited states since they entail local accumulation of vorticity and,
for this reason, a higher energy cost. It follows that AA and VV
interactions concerns more frequently interactions of identical
vortices of different VA pairs rather than pairs of identical
vortices. Interactions of identical vortices become dominating
in superconductors when an external magnetic field is switched on.
This is able to break the vortex charge neutrality and allows for
vortex arrays where the vortex density is greater than the
antivortex one.
The decreasing of the antivortex fraction is compensated by the
magnetic field that can be viewed as a macroscopic background charge.
The same effect is achieved in superfluids when the condensate
undergoes a uniform rotation \cite{TITI}.

The apparent nonquantum character of VA pairs above mentioned,
disappears as soon as one goes beyond the simplified scenery
of the mean-field picture. Indeed the VA dynamics is as complex
as the VV dynamics when any elementary effect (e.g., the
boundary effects) coupling vortices with the environment is taken
into account \cite{TITI}, \cite{DON}. This is further confirmed,
at the classical level, when the substrate effects on vortex
dynamics are inserted either via an effective potential \cite{LEGU},
or by the explicit introduction of rigid obstacles in the fluid
\cite{LUSI}. A possible, unexpected effect is that of providing
the customary continuum spectrum of the VA pair, with a discrete
character \cite{PE1}.

The purpose of this paper is to provide a complete treatment
of the vortex pair QD both for the VV case and for
the VA case (the AA case and the VV case are easily shown to be
equivalent) and of attracting the attention on the rich structure
characterizing their quantization from  the group-theoretic
viewpoint in relation to possible applications.

VQD is studied by means of the spectrum generating algebra
method \cite{ZFG} which consists essentially in identifying first 
a complete set of dynamical degrees of freedom forming a Lie
algebra (the dynamical algebra) and in constructing then the
Hilbert space of the system by exploiting the unitary irreducible
representations of the related Lie group. In this sense the vortex
system is quite interesting due to the rich variety of ways in
which the vortex coordinates can be structured so as to form a
dynamical algebra. Remarkably, the possibility of working within
different algebraic schemes does not reduces to the mere freedom
of choosing among different formal approaches trivially equivalent
to each other. Each algebraic scheme, in fact, sheds light
on some particular feature of the system.

Classical dynamics of $N$ pointlike vortices is reviewed in
Sec. II together with the symmetries characterizing their motion
described by the habitual $E(2)$-like symmetry group. The procedure
adopted for quantizing the $N$ vortex gas and hence the vortex pair
is the standard canonical scheme \cite{CHM}, \cite{PE2} concerning
the pairs of canonically conjugate coordinates assigned to the points
where the vorticity field is nonzero. Moreover, the vortex pair
dynamics is pictured through a significant geometric form
suitable for interpreting the quantum spectra.

In Sec. III the QD of two interacting vortices
is examined in the case when the topological charges, namely the
vorticities $k_1$ and $k_2$ of pointlike vortices are arbitrary.
Its integrable character is clearly manifested by the fact that the
pair Hamiltonian is a function of the Casimir operator of the
$E(2)$-like dynamical algebra.
After working out explicity both the eigenstates and the
eigenvalues of the energy spectrum, the degeneration due to
the high symmetry of the vortex pair dynamics is analyzed.
We show how a complete set of eigenfunctions can be worked
out thanks to the possibility of observing one among the numerous
constants of motion of the system. As in the case of the Landau
spectrum for the electrons acted by a transverse magnetic field,
a further quantum number describing the system symmetry must be
introduced in addition to that describing the pair energy.

In Sec. IV the dynamical algebra is assumed to be su(1,1). 
This leads to reveal unexpected symmetry properties relating
dynamics of vortex pairs with different topological charges, and
provides the framework for evaluating Berry's phase when
$k_j$ and the density $\rho$ are possibly time dependent.
Recasting instead the pair Hamiltonian in the magneticlike
form of Sec. V allows one to derive the Feynman-Onsager
condition \cite{ONS}, \cite{FO} on the $j$th topological charges
$k_j \equiv h n_j/m$, where $j= 1, 2$, and $m$ is the helium
atom mass, from pure symmetry considerations.

Finally, Sec. VI is devoted to introducing a two-boson
dynamical algebra which, in addition to give a further
insight on the specific features both of the VA case and of
the VV case, provides the correct way to approach the
pair dynamics when a disklike obstacle is present in the fluid. 
Our algebraic construction reformulates the resulting disk-pair
model in terms of a generalized angular momentum dynamics,
leads to a clear geometric picture of the pair-obstacle
interaction, provides the Hamiltonian in a form exempt from
the habitual ordering problem, and establishes the basis for
investigating the quantum processes. The interest for such a
case is founded as well on the fact that the pair scattering
phenomena can be interpreted as pinning effects on vortices due
to the impurities of superfluid substrate.

A final comment is in order as to considering generic
topological charges $k_j$, despite the standard choice used
in the literature $|n_j| =1$. The motivations are at least two.
First of all, employing arbitrary charges $k_1$ and $k_2$ involves
no formal complication, whereas it allows one to gain an interesting
general insight when relating the algebraic structures pertaining
to the VV case to those of VA case. Secondly, the emergence of vortices
with $|n_j| >1$ can take place in sufficiently excited superfluid
media as well as in close proximity of the medium boundary \cite{DON}.
Such a situation is certainly interesting since a vortex pair with
$|n_1|, |n_2| >1$ provides the basic framework in which to investigate
the exchange of $\it quanta$ of vorticity between the pair members
via quantum tunnelling processes.
\section{canonical quantization of the 2-D vortex gas dynamics}
Classical dynamics of $N$ pointlike vortices in a frictionless
fluid is described by the Hamiltonian \cite{SAF}
\be
H( {\bf{R_1}},...,{\bf{R_n}}) =
-{ {\rho} \over { 4 \pi}} \, \sum_{i \ne j}
\,\, k_i k_j \,\,
{\rm ln} \Bigl( \frac{\vert {\bf{R_i}} - {\bf{R_j}} \vert}{a}
\Bigr) \, ,
\label{HAM}
\ee
where the parameter $k_j$ represents the vorticity carried by the
jth vortex, $\rho$ is the fluid planar density, and the vector
${\bf{R_j}} = (x_j, y_j)$ describes the jth vortex position in
the 2D ambient space in terms of planar coordinates $x_j$, $y_j$.
The length $a$ represents the vortex core size that is the minimum
distance allowed between a vortex ($k_j >0$) and an antivortex
($k_j <0$) before coalescence processes take place. The Hamiltonian
equations relative to Eq. (\ref{HAM}) are standardly derived via the
Poisson brackets \cite{ARE}
\be
\{F,G \} = \sum_j { 1 \over {\rho k_j}}
\Bigl( {{\partial F}\over {\partial x_j}}
{{\partial G}\over {\partial y_j}} - 
{{\partial G}\over {\partial x_j}}
{{\partial F}\over {\partial y_j}}  \Bigr ) \quad, 
\label{PB} 
\ee
involving, in turn, the $\rho k_j$-dependent canonical brackets
\be
\{ x_i ,y_j \} = {{\delta_{ij}} \over {\rho k_j}} \quad.
\label{CB}
\ee
Then vortex coordinates can be regarded as a complete set
of canonically conjugate variables whose $\it momenta$ are
defined as $p_j = \rho k_j y_j$. Also, one can easily check
that the functions
$$
J_z =  -{{\rho} \over 2} \sum_m k_m ( x_m^2 + y_m^2 ) \, ,
$$
\be
J_x = \rho \sum_m k_m x_m
\label{AGEN} \; ,
\ee
$$
J_y = \rho \sum_m k_m y_m \, ,
$$
\be
J_{*} =  {{\rho} \over 2} 
\sum_{i \ne j}\, k_i k_j\, \bigl[ (x_i - x_j)^2 +
(y_i - y_j)^2 \bigr] \; ,
\label{CAS1}
\ee
fulfilling the equation
\be
- \rho J_{*} = 2 C J_z + J_x^2 + J_y^2 \; ,
\label{CAS2}
\ee
where $C = \rho \,\, \sum_j k_j $ is related to the total vorticity, 
are constants of motion and satisfy the classical commutators
$(a = x, y, z)$
\be
\{ J_a ,J_{*} \} = 0 ,\quad \{ H ,J_a \} = \{ H ,J_{*} \} = 0 \, ,
\label{FIN}
\ee
\be
\{ J_x ,J_y \} = C \; ,\;  
\{ J_z ,J_x \} = J_y \; ,\; \{ J_y ,J_z \} = J_x \;.
\label{CAL}
\ee
It is worth noticing that $J_x, J_y, J_z$ exhibit an
${\rm e}(2)$-like algebraic structure --hereafter we shall
denote it by ${\rm e}_*(2)$-- which is fully reached when
$C =0$, namely when the vortex total charge equals the
antivortex total charge. 

The canonical quantum description of $N$ pointlike vortices is
obtained by replacing classical commutators (3) with
\be
[ x_i ,p_j ] = \delta_{ij} i \hbar \quad,
\label{QAL}
\ee
(we have set $p_j = \rho k_j \, y_j$ in order to get commutators
in the canonical form) that furnishes the quantum version of
the algebra (7)
\be
[ J_x ,J_y ] = i \hbar C \,\,,
[ J_z ,J_x ] = i \hbar J_y \,\,, [ J_z ,J_y ] = i\hbar J_x \,\, .
\label{QALG}
\ee
In view of the fact that the constants of motion can be employed
for integrating the dynamical equation provided they are in involution,
commutators (9) involves that the many-body wave function for the
2D vortex gas is characterized at most by three macroscopic quantum
numbers two of which are, of course, $H$ and $J_*$, while the third
one can be arbitrarily chosen among $J_x$, $J_y$, and $J_z$. 

A meaningful geometric picture of the system dynamics is achieved
by means of such constants of motion when the vortex pair is
considered.  To this end it is useful to describe the vortex
pair through the new set of coordinates 
\be
x \doteq x_1 - x_2 \;\; , \quad y \doteq y_1 - y_2 \;\; ,
\label{CO1}
\ee
\be
X \doteq J_x /C  \;\; , \quad Y \doteq J_y /C \;\; ,
\label{CO2}
\ee
where $J_x =\rho (k_1 x_1 + k_2 x_2)$ and
$J_y =\rho (k_1 y_1 + k_2 y_2)$ follow from Eqs. (\ref{AGEN}),
that reduce the Casimir function (\ref{CAS1}) to the form
\be
J_{*} = \rho k_1 k_2 \, \bigl( x^2 + y^2 \bigr) \; .
\label{ARG}
\ee
Then, after expressing the coordinates $x_j$, $y_j$ as
\be
x_1 = {1 \over C}(J_x +\rho k_2 x) \, ,
y_1 = {1 \over C}(J_y +\rho k_2 y) \, ,
\ee
\be
x_2 = {1 \over C}(J_x -\rho k_1 x) \, ,
y_2 = {1 \over C}(J_y -\rho k_1 y ) \, ,
\label{CC1}
\ee
by means of Eqs. (\ref{CO1}) and (\ref{CO2}), it is quite easy to
recast $J_*$ in the two equivalent forms
\be
{{J_{*}} \over {k_1 k_2 \rho}} =
  {{C^2} \over {\rho^2 k^2_2 }}  
\Bigl[\Bigl( x_1 - {{J_x} \over C} \Bigr)^2 + 
\Bigl( y_1 - {{J_y} \over C} \Bigr)^2 \Bigr] \, ,
\label{CIR1}
\ee
\be
{{J_{*}} \over {k_1 k_2 \rho}} =
   {{C^2} \over {\rho^2 k^2_1 }} 
\Bigl[\Bigl( x_2 - {{J_x} \over C} \Bigr)^2 + 
\Bigl( y_2 - {{J_y} \over C} \Bigr)^2 \Bigr] \, ,
\label{CIR2}
\ee
which, without solving the equations of motion, identify completely
the classical orbits where the two vortices move along. Such orbits
--they are easily recognized to be circumferences-- present as a
common center the vorticity center ${\bf R}_{*}=(J_x/C , J_y/C)$
and have radii
\be
{\cal R}_1 = {{\vert k_2 \vert} \over {\vert k_1 + k_2 \vert}}
\sqrt{ {J_*} \over {k_1 k_2 \rho} }
\label{RAG}
\ee
$$
{\cal R}_2
= {{\vert k_1 \vert} \over {\vert k_1 + k_2 \vert}}
\sqrt{ {J_*} \over {k_1 k_2 \rho} }
$$
\noindent
for the vortex with charge $k_1$ and $k_2$, respectively.
${\bf R}_{*}$, ${\cal R}_1$, and ${\cal R}_2$ are manifestly
time-independent quantities in that they just depends on the
constants of motion $J_x ,\, J_y ,\,J_{*}$.

For $k_2 ,k_1 > 0$ vortices rotate in such a way that a common
straightline always join them to ${\bf R}_{*}$. The latter,
in particular, coincides with the rotation center which is situated
between the two vortices. When $k_2 \rightarrow k_1$ the
circumferences merge in one whose center is yet ${\bf R}_{*}$.
On the other hand for $k_2 \rightarrow 0$ the radius ${\cal R}_1$
vanishes so that the vortex with finite vorticity $k_1$ fall in
${\bf R}_{*}$ thus  losing any dynamical role.
In this case the weak vortex --that with $k_2 \simeq 0$--
ends up by running along on a limiting circle with radius
${\cal R}_2 \rightarrow$ const.
When $k_2 \rightarrow 0$ from negative values the dynamical
situation is almost the same except for the fact that now the
vortices stay along a rotating half-line whose extreme is attached
to ${\bf R}_{*}$.
The emergence of the full VA regime is announced when, for
$k_2\rightarrow -k_1$, the center ${\bf R}_{*}$ moves away
from the vortices and gets a larger and larger distance from them.
In the limiting case of the pair with $k_2 \equiv -k_1$ the two
vortices run along parallel straight-lines (i.e. circles with
infinitely large radii) and keep a constant relative distance.
One easily checks that ${\cal R}_1,{\cal R}_2 \rightarrow \infty$.
\section{degeneracy of the pair energy spectrum}
Quantizing the two-vortex system seems no more complex than
quantizing a simple harmonic oscillator (HO) \cite{CHM}, \cite{PE2}
even when the topological charges $k_1$, $k_2$ of the interacting
vortices are arbitrary. In this case, in fact, Hamiltonian
(\ref{HAM}) reduces to a single logarithmic term whose argument
is ${\vert{\bf{R_1}}-{\bf{R_2}}\vert}^2$, while Eq. (\ref{CAS1}),
whereby $J_{*}$ takes the form
$J_{*} ={\vert{\bf{R_1}}-{\bf{R_2}}\vert}^2/ \rho k_1k_2$,
entails $H$ written as
\be
H( {\bf{R_1}}, {\bf{R_2}}) = 
- { {\rho} \over { 4 \pi}}
\,\, k_1 k_2 \,\,
{\rm ln} \Bigl( {{J_{*}} \over { k_1 k_2 \rho a^2}} \Bigr) \quad . 
\label{PH}
\ee
This feature is specific to the two-body problem and implies
that the set of energy eigenvectors exactly coincides with
the $J_*$ spectrum. A remarkable freedom is then permitted
in selecting the remaining quantum number which labels the
degeneracy of the energy states. In fact, any operator of
the form $I = aJ_x + bJ_y + cJ_z$ fulfills the equation
$[H, I]=0$ establishing the constant of motion status of $I$.
Anyway, a deeper inspection reveals that any invariant
$I(a,b,c)$ is obtained either from $J_z$ or from $J_y$
via the transformation $I = g I_0 g^{-1}$,
$I_0 = J_y ,\, J_z$, where $g$ is a unitary
transformations obtained by combining appropriately
the action of $D_x( \alpha) = {\rm e}^{i\alpha J_x}$,
$D_y(\beta)={\rm e}^{i\beta J_y}$,
and $D_z(\phi) = {\rm e}^{i \phi J_z}$.
The possibility to reconstruct the algebra of the $E_*(2)$
group (the symmetry group of $H$) from the elements $J_z$ and
$J_y$, representative of the algebra disjoint sectors, via the
adjoint action map, entails two possible pictures of the degeneracy.
In this section we examine the vortex pair spectrum relative to two
such ways to structure the energy--level degeneracy.

To begin with we assume that the topological charges fulfill the
inequalities $k_1 > 0 ,\quad 0\le \vert k_2 \vert \le k_1$
and notice how the ranges allowed are capable of describing any
possible pair. Also, let us introduce the compact notation
${\bf D} ={\bf{R_1}} - {\bf{R_2}}$, and express it by means
of the set of canonical conjugate variables
[see Eqs. (\ref{CO1}), (\ref{CO2})]
\be
X= \frac{J_x}{C} \, , \;  x= x_1-x_2 \, , \:
p=\frac{k_1k_2\rho}{k_1+k_2}  y \, , \;
P= J_y \, ,
\label{PP1}
\ee
whose momenta $p$ and $P$ satisfy the commutators $[X,p]=[x,P]=0$
and $[x,p]=[X,P]=i \hbar$. By using such variables, the logarithm
argument is easily turned into the HO form
\be
{\bf D}^2 =  y^2 + x^2 
=  {1 \over {w^2}} \bigl(p^2 + w^2 x^2 \bigr) \; ,
\label{HOF}
\ee
where the frequency $w$ reads $w = k_1 k_2 \rho / (k_1 + k_2)$.
This implicates that the wave functions
\be
\Psi_n(x ;\vert w \vert ) = { 1 \over {\sqrt { 2^n \pi \, n! \ell}}}
{\rm e}^{ - {{x^2} \over {2 \ell^2}}} H_n( x / \ell ) \quad ,
\label{WF1}
\ee
where $\ell^2 = \hbar / \vert w \vert$, that satisfies
the secular equation $\bigl( p^2 + w^2 x^2 \bigr)\, \Psi_n
=\hbar \, \vert w \vert \,( 2n + 1 ) \Psi_n$,
represent the eigenfunctions of ${\bf D}^2$
with eigenvalues given by
\be
S_n (w ) = (\hbar / \vert w \vert )( 2n +1 ). 
\label{EV1}
\ee
A complete set of eigenfunctions $\Psi_{n K}(x ,X )$ is finally
obtained when the further observable $P = J_y$ is considered
together with ${\bf D}^2$. The second quantum number $K$ in
\be
\Psi_{n K}(x,X) \doteq \Psi_n(x) \Phi_K (X) \; ,
\label{CWF1}
\ee 
where the plane wave
$\Phi_K (X) \doteq {\rm e}^{i KX}/{\sqrt{2\pi}}$
fulfills the equation $P \Phi_K (X) = \hbar K \, \Phi_K (X)$,
establishes the position $Y = P /C$ of the center of vorticity
along the $y$ axis in the ambient space.

As an alternative, a complete set of eigenstates can be
constructed by resorting to the quantum number related to
the conserved quantity $J_z$. This is easily achieved by
exploiting Eq. (\ref{CAS2}) in that, after turning it to the
form $J_z =-(\rho J_* +J_x^2 +J_y^2)/2C$,
it is evident that the HO-like wave function
\be
\Psi_m(X ;\vert C \vert ) \equiv
{ 1 \over {\sqrt { 2^m \pi \, m! L}}}
{\rm e}^{ - {{X^2} \over {2 L^2}}} H_m( X / L ) \, ,
\label{WF3}
\ee
where $L = \sqrt{\hbar/ \vert C \vert}$, diagonalizes
$J_x^2 + J_y^2 =P^2 + C^2 X^2$. Hence the eigenvalues
associated with $J_z$ have the form
\be
\Lambda_m(n;w) = - \hbar \Bigl ( sg(w )  n + m + 1 \Bigr)
\, ,
\label{EV2}
\ee
where $sg( w ) \equiv w / \vert w \vert$ and we have exploited
the fact that $\rho J_* /C =\hbar (2n+1) w/ \vert w\vert$
and $C =\rho (k_1 + k_2) > 0$, due to our initial assumptions.

The descriptions of the energy spectrum degeneracy just
examined, involve a significant geometric-quantum picture.
Using the quantum numbers $(n, K)$ implies that the two
vortices are confined along the two circumferences
(\ref{CIR1}), (\ref{CIR2}) --the angular coordinates cannot
be specified due to the uncertainty principle-- whose radii
\be
{\cal R}_j =  
{{\vert k_1 k_2 \vert} \over {\vert k_j \vert \, \vert C \vert}}
{\sqrt{S_n (w)}} \, ,
\label{RAS}
\ee
where $j= 1, 2$, are now labelled by the integer $n$
due to Eq. (\ref{EV1}). The center of such circumferences, which
coincides with the vorticity center, has $Y = \hbar K /C$ while
$X$ is, as expected, undetermined.

On the other hand, when the pair $(n,m)$ is employed
the $\it locus$ allowed for the vorticity center changes
from noncompact to compact. The latter, in fact, is now
confined on the circle of radius
$$
X^2 + Y^2 = [\hbar (2m+1) / \vert C \vert]^{1/2} \quad,
$$
labeled by $m$, instead of a straight line labeled by $K$.
Once again the uncertainty principle prevents one from getting
any further information both on the position of the vorticity
center and on the vortex position along the circles of radii
${\cal R}_1$ and ${\cal R}_2$.

In the extreme case when $C= k_2 + k_1 \rightarrow 0$ the
canonical scheme based on Eqs. (\ref{PP1}) breaks down due
to the divergence of the factor $1/(k_1 + k_2)$. In particular,
the singular behavior of the pure VA case is distinguished by
the fact that $J_x$ and $J_y$ end up by coinciding with $x$ and
$y$, respectively, which now commute since
$\rho k_1 [x,y] \equiv [J_x ,J_y]= i \hbar C\equiv 0$.
Such circumstances impose the introduction of a more
appropriate set of canonically conjugate variables. We thus define
\be
\eta= - \rho k x \, , \;
{\cal Y}= \frac{y_1 +y_2}{2} \, , \;
\xi = \rho k y \, , \;
{\cal X}= \frac{x_1 +x_2}{2} \, ,
\label{VAV}
\ee
where $k_2 =- k_1 = -k$, which turns out to be
completely disjoint from those employed in the case
when $k_2 \ne - k_1$, and obeys the standard relations
$[{\cal X},\xi ]=[{\cal Y},\eta ]=i\hbar$. Here $\cal X$,
$\cal Y$, $\xi$ and $\eta$ do not play prefixed roles
so that, depending on the interactions involved by the
dynamical problem, they can be regarded either as
$\it momenta$ or as position variables.
When further interactions are excluded from $H$, the simplest
choice is that where $\cal X$ and $\cal Y$ are looked upon as
coordinates which implies that the energy eigenfunctions have
the form
\be
\Phi_{\bf K} ({\cal X},{\cal Y}) = { 1 \over {2 \pi}}
{\rm e}^{i({\cal X} K_{\cal X}+ {\cal Y} K_{\cal Y})} \,  .
\label{WF4}
\ee
Here $\hbar K_{\cal X}$ and $\hbar K_{\cal Y}$
are the eigenvalues of $\xi=-i\hbar \partial_{\cal X}$
and $\eta=-i\hbar \partial_{\cal Y}$, respectively.
Information on ${\cal X}$ and ${\cal Y}$ are, of
course, completely missing that is the pair cannot
localized anywhere in the ambient space.

Some applications can be now illustrated. The quantum mechanical
problem just solved provides the formal tools requested for
investigating the scattering processes of the VV pair 
(as well as of the AA pair) and VA pair dynamics in the presence of
2D potential wells simulating the confining action of the defects
placed on the superfluid medium substrate \cite{LEGU}.
The effective model Hamiltonian
\bq
{\cal H} = {g \over 4} [(x_1 + x_2)^2 + (y_1 + y_2) ^2] + H \, ,
\nonumber
\eq
where $g$ represents the strength of the phenomenological
confining action, for $k_1 =+k_2 =k$ and $k_1 =-k_2 = k$ yields
the Hamiltonian
\be
{\cal H} = - { g \over {\rho k}}  (J_z + J_* / 4k)
- { {\rho} \over { 4 \pi}}
\,\, k^2 \,\, {\rm ln}
\left [ {{J_{*}} \over { k^2 \rho a^2}} \right ]
\, ,
\label{HFV}
\ee
and
\be
{\cal H} = g( {\cal X}^2 + {\cal Y}^2)
+ { {\rho} \over { 4 \pi}} \,\, k^2 \,\,
{\rm ln}
\left [ {{\xi^2 +\eta^2} \over {(\rho k a)^2}} \right ]
\, , 
\label{BU}
\ee
respectively. In view of the spectral problems solved above,
one finds that Hamiltonian (\ref{HFV}) has a spectrum which
is readily obtained from Eqs. (\ref{EV1}) and (\ref{EV2}),
whereas Hamiltonian (\ref{BU}) is clearly related to a
2D Coulomb problem where two particles with opposite
charges are studied within the center-of-mass reference
frame and are endowed with a reduced mass
$\mu = {{\rho^2 k^2}/ {2g}}$. Notice how it is now natural to
exchange the roles of ${\cal X}$, ${\cal Y}$ and $\xi$, $\eta$
assigned above. The treatment of such a Coulomb system will
be reconsidered in Sec. VI where a more adequate algebraic
scheme will be introduced.

More in general, Hamiltonians of the form ${\cal H} = f(I) + H$,
where $f(I)$ is a generic function of the operator $I(a,b,c)$
defined above, are easily diagonalized by reducing $I$ either
to $J_z$ or to $J_y$, depending on the values taken by $a$, $b$,
and $c$. For example, the situations where $\cal H$ exhibits
a term $I$ proportional either to $J_z$, or to $J_x$ ($J_y$)
can be interpreted as the way to picture the effect of a
macroscopic velocity field responsible for a uniform
rotation around the plane origin, in the first case,
and inducing the vortex dragging along the
$y$ ($x$) axis, in the second case.

A final comment is in order as to the two limiting
cases $k_1 \rightarrow \infty$ with finite $k_2$
(the vortex with $k_1$ recover the classical status
since $[x_1, y_1]$ is vanishing), and $k_2 \rightarrow 0$
with finite $k_1$ (the vortex with $k_2$ gets an ultraquantum
status since $[x_2, y_2] \rightarrow \infty$).
At the classical level such limits make the $k_1$ vortex
tend to stillness while the $k_2$ vortex goes on to rotate
with a frequency proportional to $w \rightarrow \rho k_2$.
In particular a greater and greater period occurs when
$k_2 \rightarrow 0$. While the realism of a situation where
$k_2 \simeq 0$ is hard to maintain, since there is no experimental
evidence of fractional quanta of vorticity, the case when $k_1$
is large can be easily interpreted as the situation where
a small cluster of vortices interacts with solitary vortices.
Moreover, while the first case implies diverging spectrum
gaps $S_{n+1}-S_n$ due to Eq. (\ref{EV1}), the second one leads to
$S_n = \hbar (2n+1)/\rho k_2$.
In particular $k_1 \rightarrow \infty$ is related
to the dynamics of a solitary vortex moving around a 
disklike obstruction with radius $R$ contained inside
the 2D ambient space.
Hamiltonian $H_D$ of Appendix A embodying the effects
of the disk is illustrative of the manner in which this is
realized when the vortex  complex coordinate 
$z_1$ is such that $|z_1|  \simeq R$.
%
%
\section{the su(1,1) approach}
A different approach to the quantization of the vortex pair
is provided by the $xp$ realization of the algebra su(1,1)
which allows one to regard
${\bf D}^2 ={\vert {\bf{R_1}} - {\bf{R_2}} \vert}^2$ as the
compact generator of su(1,1). After recalling that the
algebra generators are given by
\cite{BPS} 
\bq
J_1 = {{w^2 x^2  - p^2 } \over {4\hbar \vert w \vert}} \,, \,
J_2 = - {{ p x + x p } \over {4\hbar}} \, , \, 
J_3 = {{w^2 x^2  + p^2 } \over {4\hbar \vert w \vert}} \, ,
\nonumber
\eq
and satisfy the commutation relations
\be
[J_1, J_2] = -iJ_3 \, , \, [J_2, J_3] = i J_1 
\, , \, [J_3, J_1] = iJ_2 \, ,
\label{UUC}
\ee
the equation ${\bf D}^2 \equiv 4 \hbar J_3 / \vert w \vert$
ensuing from Eq. (\ref{HOF}) shows that the spectrum of $J_3$
is that involved by the vortex dynamics. It shows as well
how ${\bf D}^2$ no longer plays the role of the Casimir operator
now being nontrivially acted both by $J_1$ and by $J_2$.
In spite of this the present algebraic framework inherits
both $X$ and $P=C Y$ (and hence any function depending on them)
as dynamical constants of motion from the ${\rm e}_*(2)$ algebra
of Sec. III.  As for Hamiltonians (\ref{HFV}), (\ref{BU}), again
one can take advantage of this fact for constructing vortex models
with Hamiltonian of the form ${\cal H} = F(X, Y) + H$ accounting for
the background medium influence. For example,  describing the drag
action on the pair vorticity center due to the flow stream lines in
the presence of a saddle point simply requires
$F(X,Y)=\gamma (X^2-Y^2)$,
$\gamma$ being some suitable dimensional parameter.

The su(1,1) scheme is useful to discuss the statistics of the
pair system.  In Ref. \cite{LEMY} the same algebra was constructed
step by step starting from the Weyl-Heisenberg algebra
$\{{\bf I}, x, p \}$, in order to relate the VV pair statistics to
the unitary irreducible representations (UIR) of the $xp$ realization.
A direct way to obtain them is that of calculating the eigenvalues
of the su(1,1)-Casimir operator
$$
C = {J_3}^2 -{J_1}^2 -{J_2}^2 = l(l+1) {\bf I}
$$
where ${\bf I}$ is the identity operator. One easily finds
that the allowed values for $l$ are $l = -1/4$ and $l = -3/4$
which select two UIR's in the set of the SU(1,1) supplementary
series. The solutions of the secular equation
$J_3 \,f_{\nu}(x ; l)\, = \nu\, f_{\nu}(x ; l)$
(see Ref. \cite{PE3}) read
$$
f_{\nu}(x; l) = (-)^s D_{ls} \Bigl({2 \over {\ell}} \Bigr)^{1/2} 
\Bigl({x \over {\ell}} \Bigr)^{\alpha + 1/2}\,
{\rm e}^{- x^2 /2 \ell^2}
L_{s}^{\alpha}(x^2 / {\ell}^2) \, ,
$$
where $\ell$ is the same dimensional parameter employed in
Eq. (\ref{WF1}), $D_{ls} = [ s! / \Gamma(s-2l)]^{1/2}$ is the
normalization factor and $L_{s}^{\alpha}$ are the Laguerre
polynomials, whereas $\alpha$ and the nonnegative integer
$s$ are related to $l$ and $\nu$ by $\alpha = - (2l + 1)$ and
$\nu = s - l$, respectively. By exploiting the
general formula \cite{ABST}
$$
L_{s}^{\alpha}(z^2) = {{ (-)^s H_{n(\alpha)}(z)} \over 
{2^{n(\alpha)}\, s! \, z^{\alpha + 1/2} }} \quad ,
$$
where $n(\alpha) = 2s + \alpha + 1/2$, relating Laguerre
polynomials to Hermite polynomials $H_{n(\alpha)}$ when
$\alpha = \pm 1/2$, one finds that
$f_{\nu}(x; l) \equiv \Psi_{n}(x ; \vert w \vert)$ namely the
functions (\ref{WF1}).
The representations corresponding to $l = -3/4$ and $l = -1/4$
are thus associated with symmetric and antisymmetric
eigenfunctions respectively, in that 
$\Psi_{n}(-x ; |w|) = (-)^n \Psi_{n}(x ; |w|)$ and
$$
f_{s+ 1/4}(x;\, -1/4) = \Psi_{2s}(x ; \vert w \vert) \, , 
$$
$$
f_{s+ 3/4}(x;\, -3/4) = \Psi_{2s + 1}(x ; \vert w \vert) \, .
$$
This establishes when the pair has either a fermionic character 
or a bosonic character with respect to the transformation
$({\bf R}_1,{\bf R}_2)\rightarrow ({\bf R}_2,{\bf R}_1)$
changing $x$, $y$ (namely $p$) in $-x$, $-y$. No conclusion,
however, is permitted until the second quantum number requested 
for the complete description of the pair, is considered. To this end
consider the state $\Psi_{nm} =\Psi_n (x ; |w|) \,\Psi_{m}(X;|C|)$
obtained from formulas (\ref{WF3}), (\ref{WF1}). When the
vortex exchange is equivalently enacted via the substitution
$(k_1, k_2)\rightarrow (k_2, k_1)$, this implicates, in particular,
that $X \rightarrow X' \equiv X + x  (k_2-k_1)/C$.  Hence, while
the charge exchange does not affect $\Psi_n (x ; |w|)$ for any
value of $k_1$, $k_2$, thus exhibiting an unexpected type of symmetry
involving nonidentical charges, the usual situation is re-established
by the presence of $\Psi_{m}(X; |C|)$ which is trivially symmetric
only when $X' =X$, i.e., $k_1 = k_2$.

A further aspect that makes interesting to adopt the
su(1,1) scheme is connected to the effect of the unitary action of
$D_{\phi}= {\rm exp}(i\phi J_2)$ on the canonical variables $x$, $p$.
In passing we point out how this is the distinctive trait of the 
$xp$ description which, as opposite to the ${\rm e}_{*}(2)$ scheme,
does not involve for $\bf D$ the role of a constant, structureless
object. The $D_{\phi}$ action is given by \cite{PE3}
$$
D_{\phi} x D_{\phi}^{\dagger} = {\rm e}^{-\phi/2} x \, , 
\;
D_{\phi} p D_{\phi}^{\dagger} = {\rm e}^{\phi/2} p 
$$
and implies that
$D_{\phi}^{\dagger}\,\Psi_n (x;r)=\Psi_n(x; r{\rm e}^{-\phi})$
for any wave function (\ref{WF1}). Equipped with such formulas,
one easily shows $D_{\phi}$ to succeed in connecting the
dynamics characterized by $(k_1, k_2)$ and
$w = k_1 k_2 \rho / (k_1 +k_2)$, with any
other having different vorticities $(K_1, K_2)$ and
$W \equiv K_1 K_2 \rho / (K_1 +K_2)$. Notice that
whenever two cases are related then they both must
stay either in the VV sector, or in the VA sector. 

To exploit the $D_{\phi}$-action effects, 
we first define the modified Schr\"odinger problem
$i \hbar \,\partial_{\tau} \Phi_{\tau} = H(W)\, \Phi_{\tau}$,
with time $\tau$, where the pair Hamiltonian
$H(W)= (-\rho K_1 K_2 /4\pi)\,  {\rm ln}
[( x^2+p^2/W^2)/a^2]$ [see Eqs. (\ref{ARG}),
(\ref{PH}), and (\ref{HOF})] depends on the vorticities $K_j$.
Then, exploiting the fact that
\be
W^2 x^2  + p^2 = {\rm e}^{-\phi}\; D_{\phi}
\,(w^2 x^2  + p^2)\, D_{\phi}^{\dagger}\;\;,
\label{DF}
\ee
where $W = w {\rm e}^{-\phi}$, and setting
$\Phi_{\tau} \equiv {\rm e}^{i \theta(\tau)} D_{\phi}\, \Psi_{t}$,
with $t= t(\tau)$,
we recast the Schr\"odinger problem in the form
$$
i \hbar \left [ i{\frac{d\theta }{d\tau}} +
D_{\phi}^{\dagger} \partial_{\tau} D_{\phi}
+\partial_{\tau}) \right] \,\Psi_t=
$$
\be
=\left [ -\rho \phi K_1 K_2 / 4 \pi +
{\frac{K_1 K_2}{k_1 k_2}} H(w) \right ] \Psi_t \;,
\label{DIL}
\ee
where $\theta$ is obtained by imposing
$\hbar d\theta/ d\tau \equiv \rho \phi K_1 K_2 / 4 \pi$. 
This, in turn, reproduces the initial problem with the
charges $k_j$
\be
i \hbar \partial_t \Psi_t = H(w)\, \Psi_t \, ,
\label{HO1}
\ee
whose solutions can be derived by means of wave functions 
(\ref{WF1}), when both the condition $d\phi/dt \equiv 0$
and the time rescaling $\tau = (k_1 k_2/ K_1 K_2) t$ 
are assumed. Therefore, replacing $w$  with $W$, which
represents a generic change $k_j \rightarrow K_j$, is
compensated by substituting $\Phi$ with its transformed
version $\Psi$.

The analysis developed shows how evaluating the Berry phase
\cite{BER} when a vortex with constant charge interacts with a
vortex cluster situated at a distance much larger than the cluster
size. The vortex with slowly varying vorticity  can be regarded
as the pointlike approximation of the cluster exhibiting vortex
creation--annihilation processes.

As to this case, suppose that the problem relative to $H(W)$
has $W$ with $K_2$ depending on the time $\tau$. Then, in the
spirit of the adiabatic approximation approach \cite{BER},
we go back to Eq. (\ref{DIL}) and solve it, with the ket notation,
by setting the following two equations:
$$
\langle \Psi_t | [i\hbar\,\partial_t -H (w)]
|\Psi_t \rangle \equiv 0 \, ,
$$
$$
\hbar {\frac{d\theta }{d \tau}} \equiv
- \hbar\, 
\langle \Psi_t | J_2 |\Psi_t \rangle \, 
{\frac {d\phi }{d\tau}}
+ {\frac{\rho}{4 \pi}} K_1 K_2(\tau) \phi(\tau)
\, ,
$$
where the fact that $\phi$ is a function of the
time-dependent charge $K_2$ provides the nonvanishing term
$D_{\phi}^{\dagger} \partial_{\tau} D_{\phi}=iJ_2 (d\phi /d \tau)$.
The first equation is obtained by absorbing the dependence on the
time $\tau$ in the time $t$ via the equation
$dt/d\tau = [ K_1 K_2(\tau)/k_1 k_2]$
where both $K_1$ and $k_j$ are independent of $\tau$.
It involves the usual time-independent pair dynamics
whose exact solutions are given by Eq. (\ref{WF1}).
On the other hand, the second equation, expressing
the standard approximation of the adiabatic scheme, explicitly
provides $\theta(\tau)$ via integration which, as expected,
plays the role of geometric contribution to Berry's phase.
It is found that
$$
| \Phi_{t(\tau)} \rangle \simeq
{\rm e}^{i [\theta(\tau) -\tau E_n / \hbar]} 
\, {\rm e}^{i \phi(\tau)  J_2 } |\Psi_n \rangle \, ,
$$
where $E_n =-(\rho/4 \pi) k_1 k_2{\rm ln}[ S_n(w) /a^2]$
(see Eq. (\ref{EV1})),  $\phi$ is determined by
$\phi= -{\rm ln}(w/W)$, and $|\Psi_n \rangle$ is given by
Eq. (\ref{WF1}).
Effects due to a possible time dependence of both the density
$\rho$ and the core size $a$ can be treated
along the same lines (see Ref. \cite{HW}).
%
%
\section{Magnetic form of two-vortex dynamics}
Hamiltonian (\ref{PH}) can be easily turned into a magneticlike
form \cite{MAG} by introducing the momenta
$P_x =\rho \vert k_1 k_2 \vert^{1/2}x$ and
$P_y =\rho \vert k_1 k_2 \vert^{1/2}y$.
Such a picture allows one to recover the Feynman-Onsager
quantization condition on the charges $k_j$ in an alternative way. 
The momenta just introduced, whose range of validity
covers both the case $k_1 >0,\, k_2 >0$ and the case 
$k_1 >0,\, k_2 <0$, lead to rewrite ${\bf D}^2$ as
\be
{\bf D}^2
= {\frac{1}{\vert k_1 k_2 \vert {\rho}^2}} (P_x^2 + P_y^2) \, ,
\label{RAM}
\ee
where $P_x$, $P_y$ obey the commutator $[P_x ,P_y]=+(-)i \hbar C$
when $k_2 > 0$ ($k_2 < 0$), and allows one to identify the total 
vorticity $C$ as the parameter playing the role of the magnetic
field. Likewise, since $[P_x,J_y]=[P_y ,J_x]=0$, and $[J_x,J_y]=iC$,
it is quite natural to regard $J_x$ and $J_y$ as the generators
of magnetic translations pertaining to the present contest. They,
in fact, generates the Euclidean transformations of the vortex
coordinates
\be
D_y(\lambda_x) \, x_i \, D_y^{\dagger}(\lambda_x) 
= x_i + \lambda_x  \, , \, 
\ee
$$
D_x^{\dagger}(\lambda_y) \, y_i \, D_x(\lambda_y)
=  y_i + \lambda_y \, , 
\label{DIS1}
$$
responsible for the displacements of the vortex pair,
where $D_y(\lambda_x) ={\rm exp}(i {\lambda_x} J_y / \hbar)$,
and $D_x(\lambda_y) ={\rm exp}(i {\lambda_y} J_x / \hbar)$.

After that one can proceed along two independent lines.  
First, one can look upon $x_1$, and $y_2$ as position variables
thus defining $P_1 \equiv \rho k_1 y_1$, $P_2 \equiv-\rho k_2 x_2$
as their canonically conjugate momenta. On the other hand, the
opposite choice, where $x_2$ and $y_1$ are position variables and 
$P_2 = \rho k_2 y_2$, $P_1 =- \rho k_1 x_1$ the respective momenta,
is equally natural. The same twofold choice characterizes
the case of a planar charge acted by a transverse magnetic field.
In fact, the momentum space picture is always allowed as an
alternative way to describe the system in the coordinate space.
The interchangeable role of the vortex variables makes
vanishing such a distinction for the vortex pair system
where the ambient space contains both the momentum space 
and the configuration space. 

Assuming now to operate within the first of the above schemes we
implement the diagonalization of ${\bf D}^2$ in the Landau gauge
\cite{MAG}.
To this end Eq. (\ref{RAM}) must be recast in the more adequate 
version depending on $x_1$, $y_2$, $P_1$ and $P_2$
\be
{\bf D}^2 =
{\frac{1}{\rho^2 k_1^2}} {\cal P}_1^2 
+ {\frac{1}{\rho^2 k_2^2}} {\cal P}_2^2 \, ,
\label{MAH}
\ee
where ${\cal P}_1=P_1 -\rho k_1 y_2$, and
${\cal P}_2 =P_2 +\rho k_2 x_1$ have been singled out so
as to fulfill the conditions $[{\cal P}_j,J_y] = [{\cal P}_j ,J_x]=0$, 
and $[{\cal P}_1,{\cal P}_2]= -iC$. 
Then, by acting on Eq. (\ref{MAH}) through the gauge transformation
${\rm exp}(i \rho k_1 x_1 y_2 / \hbar)$ which turns it into the 
standard harmonic oscillator form, the Landau gauge eigenvectors
are found to be
$$
\Phi_{n,q}( x_1, y_2)= {\rm e}^{{i \over {\hbar}}(\rho k_1 x_1 -
\hbar q) y_2 }\,\Psi_n(x_1 - \hbar q / C ; \Omega) \, ,
$$
where $\Psi_n(x ;\Omega)$ is obtained from wave function
(\ref{WF1})  when $\vert w \vert$ is substituted with
$\Omega = \vert C k_1 / k_2 \vert$, whereas the associated
eigenvalues reproduce the spectrum $S_n(w)$ defined 
by Eq. (\ref{EV1}). It is easily shown as well that $J_x$ is 
diagonalized by $\Phi_{n,q}( x_1, y_2)$ and exhibits
$\hbar q$ as eigenvalues.

Now, the invariance of $H$ under the action of both
$D_{x}(\lambda_y)$, and
$D_{y}( \lambda_x)$, $\lambda_y, \, \lambda_x \in {\bf R}$, 
can be displayed in the Hilbert space through the formulas 
$$
D_{x}(\lambda_y)\, \Phi_{n,q}( x_1, y_2)
= {\rm e}^{i q \lambda_y} \,\, \Phi_{n,q}( x_1, y_2) \;,
$$
$$
D_{y}(\lambda_x)\,
\Phi_{n,q}( x_1, y_2)= \Phi_{n,q -\alpha}( x_1, y_2)\;,
$$
where $\lambda_x = \alpha /C$, and $D_{y}$ appears to act
as a raising (lowering) operator on the quantum number $q$ if
$\alpha < 0$ ($\alpha > 0$). Hence $D_{y}$ is able to explore the
range of the energy spectrum degeneracy related to the $n$th
Landau-like level.

The final step of the magnetic procedure consists in stating the
flux quantization as a consequence of imposing the expression
\be
D_{x}(\lambda_y)\,D_{y}(\lambda_x)\,
D_{x}^{\dagger}(\lambda_y)\,
D_{y}^{\dagger}(\lambda_x)\, = \,{\bf I} \,
{\rm exp} \Bigl[ 
{{i \lambda_x \lambda_y} \over { \hbar}} C  \Bigr]
\label{A5}
\ee
--it represents the displacement of the charge along a rectangular
loop on the plane $\{(x_1 , y_2)\}$-- to reduce to the $\it identity$
operator $\bf I$. The right-hand side of Eq. (\ref{A5}) is carried out
by means of the Baker-Campbell-Hausdorff formula \cite{ZFG}
${\sl e}^{a + b} = {\sl e}^{-{1 \over 2} [a,b]}\;
{\sl e}^a \; {\sl e}^b$
and becomes $\bf I$ if
\be
{{\rho \lambda_x \lambda_y} \over h} (k_1 + k_2) \equiv
{{M_{tot}} \over h} (k_1 + k_2) = N_*
\label{A6}
\ee
with $N_* \in {\bf Z}$. The quantity $M_{tot}$ is the superfluid
mass enclosed in the box of area $\lambda_x \lambda_y$.
Upon introducing the helium atomic mass $m_H = M_{tot} /N_A$,
one derives from Eq. (\ref{A6}) the constrain 
\be
k_1 + k_2 = (N_* / N_A) {h \over m_H} \quad
\label{A7}
\ee
on the pair total vorticity. Such a result can be readily extended 
to a many-vortex system in that the translation symmetry holds
independently from the number of interacting vortices considered,
as follows from (the quantum version of) Eqs. (\ref{AGEN}) and
(\ref{FIN}). Since circulation operator (\ref{A5}) only depends on
the algebraic properties of symmetry generators, the extension
is simply performed by replacing $k_1 + k_2$ with $\sum_j k_j$
in the previous formula.

A first interpretation of $N_*$ follows from Eq. (\ref{A7}) when
the Feynman-Onsager condition on the vorticity quantization is taken
into account \cite{ONS}, \cite{FO}. In fact, assuming 
$k_j = h n_j/ m_H$ with $n_j \in \bf Z$ successfully solves Eq. 
(\ref{A7}) and implies that $N_* = N_A\, \sum_j n_j$.
On the other hand,  Eq. (\ref{A7}) naturally contemplates the
Feynamn-Onsager condition as a possible solution which, in
conclusion,  appears to emerge as a pure consequence of the
symmetries characterizing the vortex system.  

Further information concerning the meaning of $N_*$ is
obtained when considering the system in a rectangular box.
A standard requirement consists in enforcing the cylinderlike
geometry in the ambient space via the further condition
$D_{x}(\lambda_y) = \bf I$ on the $y$ translation symmetry,
where $\lambda_y$ has been identified with one of the two
macroscopic dimensions characterising a 2D superfluid sample
of area $\lambda_y\lambda_x$. In the Hilbert space, this
amounts to stating the quantization condition
$q = 2 \pi s / \lambda_y$ with $s \in {\bf Z}$, 
involving the periodicity condition
$D_{x}(\lambda_y) \Phi_{n,q}(x_1, y_2)= \Phi_{n,q}(x_1, y_2)$.

We have thus recovered for the vortex pair dynamics the
description in terms of Landau levels and of their degeneracy:
$s$ enumerates the straight lines (parallel to the $y$ axis)
characterized by the fact that $X = J_x/C =$ const representing
the 1D domains of the ambient space where the vorticity center
is allowed to stay. Moreover, assuming that the eigenvalues
$\hbar q /C$ of $X\equiv J_x/C$ take values inside
$[0,\lambda_x]$ entails 
$$
0 < q \le C \lambda_x \lambda_y / h = N_* \;, 
$$
so that $N_*$ turns out to be the parameter measuring the
degeneracy as in the magnetic case.  In the case when
two equal vortices occupy an area $\lambda_x \lambda_y$
having a macroscopic size, then the degeneracy $N_* = 2 N_A$
is macroscopically large since $N_A$ is the number of atoms
contained inside that area.

As expected, in view of the analysis of the VA case developed
in Section III, when $C= k_1 + k_2 \rightarrow 0$ no magnetic
scenery can be realized this reflecting the fact that $[J_x,J_y]=0$
and the impossibility to relate finite areas of the plane with the
free-particle character of the VA dynamics quantum states.   
%
%
%
%
\section{vortex pair interaction with disklike obstruction}
Another way to quantize the vortex pair dynamics is that based
on expressing ${\bf D}^2 ={\vert {\bf{R_1}}-{\bf{R_2}} \vert}^2$
by the two-particle operator realizations either of the algebra
su(2), or of algebra su(1,1). This requires that vortices
are considered as individual objects and involves the use of
commutators (\ref{QAL}) for $x_j$ and $p_j = \rho k_j y_j$.
The purpose of this section is to show how such an approach is
particularly suitable to deal with the case when the vortex
dynamics takes place in the presence of circular obstacles with
reflecting walls.

To begin with we consider the VV dynamics, where $k_1 , k_2 >0$, 
and show its version in terms of two-particle generators
of su(2). These, when expressed via canonical
variables $x_j$, $p_j$, have the form
\be
V_3 = {1\over 4\hbar} 
\left ( r_1  p_1^2  - r_2  p_2^2 +
{{x_1^2} \over {r_1}} - {{x_2^2} \over {r_2}} \right ) \, ,
\label{AG3}
\ee
\be
V_1={1 \over 2\hbar} \left ( {\sqrt {r_1 r_2}} p_1 p_2 
+ {{ x_1 x_2} \over {\sqrt {r_1 r_2}}} \right ) \, , 
\label{AG1}
\ee
\be
V_2 = {1\over 2 \hbar}
\left ( {\sqrt  {{r_2}\over {r_1}} } x_1 p_2 - 
{\sqrt  {{r_1}\over {r_2}} } x_2 p_1 \right ) \, ,
\label{AG2}
\ee
where $r_j \equiv 1/\rho k_j$, and fulfil the standard
commutation relations of su(2)
\be
[V_1, V_2] = iV_3 \,, \; [V_2, V_3] = iV_1 \,,
\; [V_3, V_1] = iV_2 \, ,
\label{QVV}
\ee
whereas their Casimir operator is given by
$V_0 \doteq V_3^2 + V_2^2 + V_1^2 \equiv V_4^2 -1/4$, with
\be
V_4 = {1\over 4\hbar} 
\left ( r_1  p_1^2  + r_2  p_2^2 +
{{x_1^2} \over {r_1}} + {{x_2^2} \over {r_2}} \right )
\, .
\label{AG4}
\ee
Then, the fact that ${\bf D}^2$ can be expressed as
\be
{\bf D}^2 = 4 \hbar \Bigl[ {{r_1 + r_2} \over 2} \, V_4 -
{{r_2 - r_1} \over 2} \, V_3 
- {\sqrt { r_1 r_2}} V_1 \Bigr]\, ,
\label{DVV}
\ee
makes it possible to reformulate the VV dynamics within the 
su(2) scheme where the dynamical variables are now
represented by the angular momentum components
$V_j$ just defined.

The choice of the algebra su(1,1) instead characterizes the
VA case [($k_1 >0$, $k_2 < 0$)] and accounts for the change
of sign of $k_2$. Its generators $A_1$, $A_2$, and $A_3$ fulfill
the standard commutators of su(1,1)
\be
[A_1, A_2] =-iA_3\, ,\,[A_2, A_3] = iA_1\, ,
\,[A_3, A_1] =iA_2 \,,
\label{QVA}
\ee
and the equation for the Casimir operator
$A_0\doteq A_3^2-A_2^2-A_1^2 \equiv A_4^2 -1/4 $.
In particular, the explicit form of $ A_4 $, and $ A_3$ is
achieved by the substitution $r_2 \rightarrow -r_2$ in $V_3$,
and $V_4$, respectively, while $A_1$ and $A_2$ are derived by
replacing $p_2 $ with $-p_2$ in $V_1$ and $V_2$, respectively.
The operators thus obtained --notice that such substitutions can
be recast in terms of a process of analytic continuation connecting
su(2) to su(1,1))-- allows one to express ${\bf D}^2 $ as
\be
{\bf D}^2 = 4 \hbar \Bigl[  {{r_1 - r_2} \over 2} \, A_4 
+ {{r_1 + r_2} \over 2} \, A_3  - {\sqrt { r_1 r_2}} A_1 \Bigr]\;,
\label{DAR}
\ee
where $r_j\equiv 1/\rho |k_j|$. The linear character of both
Eqs. (\ref{DVV}) and (\ref{DAR}) allows one to readily obtain the
diagonal form of ${\bf D}^2$ by means of unitary transformations. 
Evidence of this is expressed by means of the formulas
\be
{\bf D}^2 =
4 \hbar \,\,  R_{\pm}
\Bigl[ {{r_1 + r_2} \over 2} \, V_4 \pm
{{r_1 + r_2 } \over 2} \, V_3 
\Bigr] R^{+}_{\pm} \;,
\label{DIV}
\ee
where $R_{\pm} \doteq {\rm e}^{i( \beta \pm \pi /2) V_2}$, 
with $ tg \beta = (r_2-r_1 )/ \sqrt{4 r_1 r_2}$, and
\be
{\bf D}^2 = 4 \hbar \,\, R_{\eta} 
\Bigl[ {{r_1 - r_2} \over 2} \, A_4 +
{{r_2 - r_1} \over 2 } \, A_3  \Bigr] R_{\eta}^{+} \;,
\label{DIA}
\ee
where $ R_{\eta} \doteq {\rm e}^{ i \eta A_2}$ and
$th \eta = \sqrt{4 r_1 r_2}/ (r_1 + r_2)$, whose basic
feature is that of depending only on $V_4$, $V_3$, and $A_4$,
$A_3$. As a consequence of the fact that $[V_4, V_j]=0$ and
$[A_4, A_j]= 0$ since (the eigenvalues of) $V_4$ and $A_4$ are
$c$ numbers labeling the representations of the respective groups,
then the spectral problem is reduced to the standard one of
diagonalizing $A_3$ and $V_3$.

The form of the vortex dynamics suggested by formulas (\ref{DVV}) and
(\ref{DAR}) is that of two interacting oscillators. In particular,
the terms $V_3$, $V_4$ and $A_3$, $A_4$ describe two independent
harmonic oscillators on their elliptic trajectories in the phase
space;  in view of the double nature of vortex coordinates pointed
out in Sec. V, such trajectories also represent circular solitary
motions of vortices around the ambient plane origin. Due to the
presence of $V_1$ and $A_1$ which introduce vortex interactions,
a more structured dynamics takes place which involves the bounded
states classically described in Sec. III.

In spite of the similarity of their diagonalization process
a crucial difference, however, distinguishes the VA case from
the VV case. The latter, in fact, presents two diagonal forms of
${\bf D}^2$, for any value of $k_1$ and $k_2$, related to the
sign $(\pm)$ of Eq. (\ref{DIV}), whereas the VA case always
exhibits a unique diagonal form of ${\bf D}^2$. Since 
\be
{1 \over 2} (r_1 + r_2 ) ( V_4  \, \pm  \, V_3 ) =
\cases{ 
&${{{r_1 + r_2} \over {4 \hbar}} ( r_1 p_1^2 + x_1^2 /r_1) }$ \cr 
& ${\-}$ \cr 
&${{{r_1 + r_2} \over {4 \hbar}} ( r_2 p_2^2 + x_2^2 /r_2)}$ \cr}
\label{DV1}
\ee
where the first and the second expressions correspond
to the case with $(+)$ and $(-)$, respectively, then the
eigenfunctions exhibit the two independent versions
\be
\Psi_{nq}^{+} (x_1, x_2) 
= R_{+} \, 
\Psi_n (x_1 ; r_1)
\,\Psi_q (x_2 ; r_2) \; ,
\label{EF1}
\ee
\be
\Psi_{nq}^{-} (x_1, x_2) 
= R_{-}
\, \Psi_n (x_2 ; r_2) 
\,\Psi_q (x_1 ; r_1)\, ,
\label{EF2}
\ee
where the wave functions $\Psi_p(x_j, r_j)$, with $p= n,q$,
diagonalize the (harmonic oscillator) secular equation
$(r_j p_j^2 + x_j^2/r_j)\,\Psi_p = \hbar (2p+1)\,\Psi_p $.
The effect of the action of $R_{\pm}$ in Eqs. (\ref{EF1}),
(\ref{EF2}) and hence the explicit form of
$\Psi_{nq}^{\pm}(x_1,x_2)$ are calculated
explicitly in Appendix B. Their relation with the
classical single vortex picture based on Eqs. (\ref{CIR1}),
(\ref{CIR2}) can be established by noticing that the circle
coordinates identify with the coordinates exhibited by the
final explicit form of Eqs. (\ref{EF1}), (\ref{EF2}). 

While $\Psi_{nq}^{\pm}(x_1,x_2)$
appear to have a form differing from that of the eigenstates
found within different procedures, the eigenvalues of
${\bf D}^2$ and $V_4 $ given by
\be
S_n=\,\hbar(r_1 +r_2)(2n+1) \; , \;
\Lambda_q(n)= {\frac{1}{2}}(n+q+1)\;,
\label{AUT}
\ee
respectively, where $n \in {\bf N}$ and $q\in {\bf N}$, are
consistent with Eqs. (\ref{EV1}) and (observe that $V_4=-J_z/2 \hbar$)
(\ref{EV2}), respectively. 

One should observe how the form of both Eqs. (\ref{EF1}) and
(\ref{EF2}) ensues from the choice of employing the
eigenvalues of $V_3 +V_4$ ($V_4 -V_3$) for describing the
degeneracy of the $n$ th level, when the diagonal form of
${\bf D}^2$ is $V_4 -V_3$ ($V_4 +V_3$). Actually a full
arbitrariness affects the previous choice since any function
$F(x_2, p_2)$ ($F(x_1, p_1)$) commutes with $V_3 +V_4$
($V_4 -V_3$). This is consistent with the scenarios
encountered in Secs. III, and IV. 

Coming now to the VA case, its distinctive feature is that
of providing, when $r_2 \rightarrow r_1$, a limiting situation
where the unitary transformation of Eq. (\ref{DIA}) is not able
to take $A_3 -A_1$ into $A_3$, such operators pertaining to
disjoint sectors of su(1,1). This consistently matches the
fact that, while $A_3$ is endowed with a discrete spectrum,
$A_3 -A_1$ exhibits instead the continuous spectrum characterizing
noncompact generators \cite{PE3}. The dramatic change
of the spectrum occurring when $k_2 \rightarrow -k_1$ is
the consequence of the transition from the regime of confinement
[described at the classical level in Sec. II by circumferences
(\ref{CIR1}), (\ref{CIR2})] to a situation where the VA pair
freely drifts through the ambient plane.

The standard, compact sector of the VA spectrum can be
readily worked out from the diagonal core of Eq. (\ref{DIA})
$A_3-A_4= (r_2p_2^2+x^2_2/r_2)/2\hbar$. It is found that
the eigenvalues and the eigenvectors of ${\bf D}^2$ are given by
$$
S_n = \hbar (r_2- r_1)(2n +1)\;,
$$
$$
\Psi_{n q}(x_1,x_2)= R_{\eta}\, \Psi_n (x_2 ; r_2) 
\,\Psi_q (x_1 ; r_1)\, ,
$$
respectively, while eigenvalues (\ref{EV2}) once more are matched
by the $A_4$ spectrum involved by states $\Psi_{n q}(x_1,x_2)$.

For $r_1 = r_2$, the eigenstates of $A_3 - A_1$ might be
easily expressed by means of the plane--wave eigenstate
of Sec. III which, however, fails in diagonalizing  the
constant of motion $A_4$. To fulfill such a requirement
one must consider functions of variables (\ref{VAV})
depending, in particular, on $\cal X$ and ${\cal Y}$.
Then ${\bf D}^2$ can be viewed as a 2D Laplacian
operator ${\bf D}^2 = (1/k\rho)^2 (\xi^2 + \eta^2) =
-({\hbar/ k\rho})^{2} (\partial_{\cal X}^2 +\partial_{\cal Y}^2)$
whose eigenvectors, in the su(1,1) setup, are given by
the Lindblad-Nagel states \cite{LINA}, \cite{BPS}
\be
\Psi_{\epsilon s}({\cal R}, \theta)= {\rm e}^{i s \theta}
J_s({\cal R}\sqrt{8\epsilon/r\hbar }) \;,
\label{LNS}
\ee
where ${\cal R}^2={\cal X}^2 +{\cal Y}^2$,
${\rm tg}\theta ={\cal Y}/{\cal X}$, and the parameter
$\epsilon \ge 0$ is the continuous eigenvalue of the secular
equation
\be
(A_3-A_1)\,\Psi_{\epsilon s}({\cal R},\theta)= 
\epsilon \,\Psi_{\epsilon s}({\cal R},\theta) \; ,
\label{EV5}
\ee
where $ A_3-A_1\equiv (r/4\hbar) (\xi^2+\eta^2)$
(see Appendix C). Concerning the index $s$, formulas
of Appendix C, besides showing the form of operators
$A_j$ in terms of variables (\ref{VAV}), allow one to 
recognize ${\rm e}^{is\theta}$ as the factor of 
$\Psi_{\epsilon s}({\cal R},\theta)$ responsible for
associating the eigenvalue $\lambda_s= -s/2$ with $A_4$.
Also, according to the Casimir formula
$A_0 =A_4^2-1/4\equiv J(J+1)$, the index $J$ labeling
the SU(1,1) turns out to have the form
$J \equiv -(|s|+1)/2$ involving negative integer or
half-integer values \cite{LINA} when $s$ takes integer values.

The geometric meaning of the eigenvalue $\lambda_s $
deserves some comment. When external interactions do
not affect the VA dynamics the pair proceeds along a
rectilinear trajectory which is orthogonal to the 
vector $\bf D$ joining its two vortices. Then the quantity
$4r \hbar A_4 = {\bf B} {\cdot {\bf D}= D B {\rm cos} \beta}$, 
where the vector ${\bf B}\doteq {\frac 1 2}({\bf R}_1+{\bf R}_2)$ 
represents the position of the pair on the plane, allows one
to evaluate the deviation
$|{\bf B}\wedge {\bf D}|/D=[ B^2 -(4\hbar rA_4/D)^2]^{1/2}$
of the pair trajectory from the plane origin.
Whenever the two vortices have the same distance from the
plane origin then $A_4=0$ since $\beta = \pm \pi/2$, and
the pair trajectory crosses the plane origin.

The algebraic approach based on the two-particle representation
just discussed is involved by the model Hamiltonian $H_*$
describing the vortex pair dynamics in the presence of a circular
boundary with radius $R$. The general form of $H_*$ where
vortices have arbitrary charges $k_1$, $k_2$ is given in
Appendix A, whereas the two extreme  cases $k_1=-k_2=k$
and $k_1=k_2=k$ are described by 
\be
H_*(z_1, z_2) = {{\rho} \over { 2\pi}} \,
\, k^2 \, {\rm ln} {\cal A}
\label{DIS}
\ee
where the vortex position vectors ${\bf R}_j$ have been
replaced with the complex variables $z_j \doteq x_j +i y_j$,
and $\cal A$ reads
$$
{\cal A}= 
R^{2u-4}(\vert z_1 \vert^2 -R^2)(\vert z_2 \vert^2 -R^2)
{\frac{\vert z_1- z_2 \vert^{2u}}
{\vert z_1 {\bar z}_2-R^2\vert^{2u}}}
$$
with $u=1$ and $u=-1$ in the case VA and VV, respectively.

Indeed the dynamics relative to $H_*$ can be profitably represented 
within the angular momentum picture introduced above,  when in the
VA (VV) case the logarithm argument is expressed as a function
${\cal A}(\{ A_j \})$ (${\cal A}(\{V_j \})$) of a 3D vector $\bf A$
($\bf V$) (see Appendix A). The first consequence is that of
simplifying the description of the system whose dynamical equations
are now obtained via the classic (up to the standard factor $i\hbar$)
commutators (\ref{QVV}) and (\ref{QVA}).  Also, the integrable
character of the dynamics with the disk-pair interaction turns out
to be accounted by a constant of motion ($A_4$ or $V_4$, depending
on the case studied) which is geometrically meaningful.
A second, {\it a priori} less evident, effect comes out within the
quantization process, where a desirable result is that of producing
operators unaffected by ordering problems.  Indeed this is the case
when, based on the angular momentum description and in view of the
logarithm property ${\rm ln} {\cal A}\equiv -{\rm ln} {\cal A}^{-1}$,
the logarithm argument of Eq. (\ref{DIS}) is reduced to
\be
{\cal A}^{-1}= {{\nu/2}\over{A_3-A_1}} + 
{{\nu^2}\over
{(A_3- \nu)^2-A_4^2}}\; ,
\label{PA1}
\ee
and
\be
{\cal A}^{-1}=
{{\nu^2}\over{ v^2 - V_3^2}}-
{{\nu^2}\over{v^2-2\nu (V_1-V_4)- V_3^2}} \; ,
\label{PA2}
\ee
where $\nu \doteq R^2/2\hbar r$ and $v \doteq V_4-\nu$.
On this account formulas (\ref{PA1}), (\ref{PA2}) provide the
most convenient way to construct the operator version
of $H_*$. A similar situation already occurred in Ref. \cite{PE1},
where the Hamiltonian of a pair interacting with a rectilinear
boundary turned out to possess at least one version able
to avoid ordering problems after the quantization process.
This fact strongly suggests that some nontrivial,  hidden
character pertains to the system presently considered as
well as to the one discussed in Ref. \cite{PE1}. As to this point
the main indication is certainly that relative to the
surviving of a constant of motion, despite the analytic
complexity induced by the presence of the boundary.

The extension of the work involved by the diagonalization
process of ${\cal A}^{-1}$ requires a separate treatment
that we shall develop elsewhere. Nevertheless, based on
the spectral analysis of the free vortex pair performed
above, it is possible to evaluate perturbatively the spectrum
of the pair dynamics when the vortex separation is much 
smaller than the distance from the disk, namely the condition
${\bf D}^2 \ll |z_j|^2$ holds for true. At the classical level,
this implies that $V_3^2,\, V_4-V_1 \ll V_4^2$, entailing
quantically the condition $S_n \ll \Lambda_q(n)$ on
eigenvalues (\ref{AUT}). Then one easily finds that
\be
\langle {\cal A}^{-1} \rangle \simeq
{\frac{\nu^3 (2n+1)}{[\Lambda_q(n)+\nu]^4}}
\Bigl[1+{\frac{\langle V_3^2 \rangle
-\nu(n+1/2)}{[\Lambda_q(n)+\nu]^2}} \Bigr] \; ,
\ee
where the expectation value notation is referred to the state 
$\Psi^{\pm}_{nq}$ in the ket notation $|n,q \rangle$.

On the other hand, in the VA case, one has
$ A_3-A_1 \, , A_4^2 \ll A_3^2$ so that the second
term of (\ref{PA1}) can be treated as a perturbation. 
When this is rewritten as
$ {\cal C}^{-1}= [(A_3-\nu-A_4)^{-1}-(A_3-\nu+A_4)^{-1}]/(2A_4)$
with ${\cal C} \equiv (A_3 - \nu)^2 - A_4^2$,
thanks to the fact that $[A_4,A_j]=0$, and the further
condition $x^2+y^2 \ll R^2 \ll {\cal X}^2 +{\cal Y}^2 $
is assumed, then it is possible to work out (see the
formulas of Appendix C) the zero--order approximation
$$
\frac{1}{\cal C}
\simeq \frac{\hbar r}{A_4} \Bigl [
\frac{1}{{\cal R}^2- u^2_+} - \frac{1}{{\cal R}^2- u^2_-} \Bigr]
\, ,
$$
where ${\cal R}^2={\cal X}^2+{\cal Y}^2$, and
$u^2_{\pm}= R^2 \pm 2\hbar r A_4$.  By using states
(\ref{LNS}) its expectation value can be expressed as
$$
{\cal I}(\epsilon,s) \equiv \langle {\frac{1}{\cal C}} \rangle \simeq
-\sum_{\delta=\pm} \, \int_{0}^{\infty}
{ \frac{2r\hbar \, J_s^2({\cal R} \mu_{\epsilon})
\,{\cal R}d{\cal R}} { \delta (u^2_{\delta} - {\cal R}^2 )}} \; ,
$$
where $\mu^2_{\epsilon} \equiv 8\epsilon /(r\hbar)$, and the
term $A_4$ in $u_{\delta}$ has been replaced by its eigenvalue
$\lambda_s =-s/2$, namely the second quantum number of states
(\ref{LNS}).  It results that
\be
\langle {\cal A}^{-1} \rangle \simeq
{\frac{\nu^2}{2\epsilon}} + 
2r\hbar \sum_{\delta=\pm} {\frac{\pi}{2}} s
J_s(\mu_{\epsilon} u_{\delta}) 
N_s(\mu_{\epsilon} u_{\delta})\, ,
\ee
where $J_s$ and $N_s$ are Bessel functions of the first and second
type, respectively \cite{GRA}.
%
%
%
%
\section{Conclusions}
In the present paper we have considered various algebraic schemes
as independent frameworks where is possible to treat the quantum
dynamics of the vortex pair both for the VV case and for the VA case.
>From the physical viewpoint, their equivalence has been checked
by showing how, for each approach, the diagonalization process
leads to the same energy spectrum. The aspect concerning the
spectrum degeneracy (scarcely mentioned in the literature and,
to our knowledge, never studied thoroughly) has been
particularly deepened. 

The introduction of a second quantum number enumerating
the degenerate states has been discussed in Sec. III, where
the vortex distance ${\bf D} =|{\bf R_1} -{\bf R_2}|^2$ (i.e.,
the logarithm argument of Hamiltonian (\ref{PH})) is identified
with the Casimir operator of the symmetry algebra ${\bf e}_*(2)$.
Any element $I(a,b,c)$ of ${\bf e}_*(2)$ can be used for describing
the degeneracy since $[I, H]=0$, although the (unitarily) independent
choices are just two, namely $I=J_y$ and $I= J_z$. The Hilbert space
basis relative to both $J_y$ and $J_z$ have been provided explicitly
and the geometric structure in the ambient space of the corresponding
spectra has been discussed in detail. These appear to mimic the stripe
structure of the Landau gauge and the Corbino disk structure of the
symmetric gauge, respectively, for a charge acted by a transversal
magnetic field. 

The parallel with the magnetically acted charge has been
completely developed in Sec. V. The Feynman-Onsager
quantization of the vortex charges is in fact  reconstructed
first by making explicit the magnetic form of $\bf D$ in terms
of generalized momenta ${\cal  P}_1$ and ${ \cal P}_2  $ (the
magnetic field is represented by the total vorticity $C$), then
by using the standard symmetry arguments leading to the
magnetic flux quantization.   

The reduction of $\bf D$ to the harmonic oscillator Hamiltonian 
with the canonical variables $x$ and $p$ has inspired instead
the use of the su(1,1) scheme of Sec. IV, where $\bf D$ no
longer plays the role of the Casimir operator.  Action (\ref{DF}) of
the noncompact generator $J_2$ on $\bf D$, which identifies with
the compact su(1,1) generator $J_3$,  allows one to recognize
a continuous symmetry relating Hamiltonian with different
vorticity pairs $(k_1, k_2)$. In view of such a symmetry
it is possible to calculate both the wave function and
the geometric correction of the Berry phase when the
Hamiltonian has time-dependent parameters. 

In Sec. VI $\bf D$ has been rewritten by means of
two-particle realizations of su(2) and su(1,1) expressed
in terms of canonical coordinates $x_j$, $p_j$ for the VV case
and the VA case, respectively. In this contest the energy degeneracy
is accounted for by the operators $A_4$ and $V_4$ labeling the algebra
representation. Such a scheme appears to be really suitable for the
purpose of treating the dynamics of vortex pairs in the presence
of a disklike obstruction in that it allows one to express the
pair Hamiltonian within an angular momentum picture, which is both
quite compact and capable of avoiding the ordering problems
despite the analytic complexity introduced by the vortex-obstruction
interaction.  The dynamics of such a case --here the energy spectrum
has been evaluated just perturbatively-- deserves further
investigations since it provides a nonphenomenological
approach to study pinning effects due to the impurities
of the medium.

More in general, the possibility of using various
algebraic approaches to treat QVD fully displays
its importance when considering, as the natural
development of the results achieved here, the constructions of
models coupling vortices with the environment, namely external
systems such as the superfluid background,  the walls confining
the superfluid, defects responsible for vortex scattering, thermal
excitations, and so on.  

In passing, we notice once more how the quantum number describing
the degeneracy could play a relevant role as the dynamical
variable to be activated by the interactions with the environment. 
In this sense it is quite natural to expect that such a number is
involved in the energy spectrum of the coupled system thus eliminating
the energy spectrum degeneracy. 

Several ways to realize the coupling with the environment
can be established depending on the algebraic framework where
the vortex dynamics is accounted. The resulting coupled dynamics
should be sensitively conditioned by the choice performed.
In particular, some scheme might appear, due to its intrisic
features, more appropriate than another one depending on
the physical contest where it is employed [this is the case,
f. i.,  of the ${\bf e}_*(2)$ scheme which clearly turns
out to be not adequate for describing the interaction with the
disklike obstruction].

A similar situation was discussed in \cite{PERA} for a charge acted
by a transversal magnetic field and interacting with the background
phonons. Indeed in that case the choice of a certain particular
dynamical algebra for the charge Hamiltonian was able to endow
the coupled model with the chaotic character requested by the
experimental observations. 

A further reason
for the interest in considering independent
algebraic descriptions of vortex dynamics is related to a possible
implementation of the time-dependent variational principle
procedure for many-vortex systems within a quantum picture
based on a coherent state picture\cite{ZFG}.
The combination of such methods has been successfully employed 
to investigate the quantum dynamics of many-body systems
\cite{MOPE}.  The main ingredient of such a variational procedure
is represented by a macroscopic wave function for the ensemble
particles whose construction is profitably realized in terms of
generalized coherent states.  
The usefulness of the analysis developed here becomes
evident by recalling that the definition of such states is  basically
founded on identifying a suitable algebraic framework (the dynamical
algebra defined in the Introduction) containing the dynamical degrees
of freedom of the ensemble.

Indeed we believe that the analysis performed in the present
work can be fruitfully employed for constructing models for the
vortex-environment interaction as well as for treating the quantum
dynamics of a gas of vortices.
%
\vskip 0.5 truecm
\centerline{\bf ACKNOWLEDGEMENTS}
\vskip 0.5 truecm
\noindent
The main part of this work was performed while the author was
visiting the International Center for Theoretical Physics (I.C.T.P.)
in Trieste, Italy. The author is grateful to the Condensed Matter
Section of I.C.T.P. for its hospitality and financial support.
\vskip 0.5 truecm
\centerline{\bf APPENDIX A}
\vskip 0.5 truecm
The standard way to account for the presence of boundaries
confining the medium where vortices move, is based
on the virtual charge method \cite{MOFE}, \cite{PE1}, \cite{LUSI}.
Such a method makes it possible to work out the vortex pair
Hamiltonian incorporating the effects of a circular boundary,
which reads
$$
H(z_1, z_2) = -{{\rho} \over { 2\pi}} \,
\, k_1 k_2 \,{\rm ln} 
\left [{\frac{ R^2 \vert z_1 - z_2 \vert^2}
{\vert z_1 {\bar z}_2 - R^2 \vert^2}} \right ]\; +
$$
$$
+ {{\rho} \over { 2\pi}} \, k_1^2 \,
{\rm ln} \left (
{\frac{ \vert z_1 \vert^2 -R^2}{R^2}} \right )
+ {{\rho} \over { 2\pi}} \, k_2^2 \,
{\rm ln} \left ({\frac{ \vert z_2 \vert^2 -R^2}{R^2}}\right )
\;.
$$
When one of the two vortices touches the disk boundary ({\it i.e.}, 
$|z_j|\rightarrow R$) then the first term is going to zero, whereas
${\rm ln}[(|z_j|^2-R^2)/R^2]$ becomes infinitely negative, as it is 
expected whenever a vortex annihilates a vortex with an
opposite charge. The latter, in the present case, is the virtual
vortex accounting for the boundary effect. Then, after performing
a suitable energy rescaling, the remaining logarithm represents
the interaction of an isolated vortex with the circular
reflecting wall.

In the extreme cases $k_1 = k_2$ and $k_2=-k_1$ the
Hamiltonian reduces to the form (\ref{DIS}).
Since $\cal A$ is constituted by several factors where the
canonical variables appear to be mixed in a very complex way,
a dramatic ordering problem should affect the quantum version
of $\cal A$. It is almost surprising instead to discover that
${\cal A}^{-1}$ is exempt from such a problem. One finds,
in fact, the expressions
$$
{\cal A}^{-1}= {\frac{R^2}{\vert z_1 - z_2 \vert^2 }} +
{\frac{R^4}{(\vert z_1 \vert^2 -R^2)(\vert z_2 \vert^2 -R^2)}}
\;,
$$
$$
{\cal A}^{-1}= 
{\frac{R^4}{(\vert z_1 \vert^2 -R^2)(\vert z_2 \vert^2 -R^2)}}
-{\frac{R^4}{\vert z_1 {\bar z}_2 - R^2 \vert^2 }}
\;,
$$
in the VA case and the VV case, respectively, that, after
considering the further formulas
$$
\vert z_1 - z_2 \vert^2=
\cases{ &4$ \hbar r(A_3 -A_1) $\cr
        & ${\-}$ \cr
        &4$ \hbar r(V_4 -V_1) $  \cr} \; ,
$$
$$
\vert z_1 {\bar z}_2 - R^2 \vert^2=
\cases{ &{\-} $ 4 \hbar^2 r^2 (A_3^2 - A_4^2) +
              R^4 -4 \hbar r R^2 A_1$\cr
        & ${\-}$ \cr
        &{\-}$ 4 \hbar^2 r^2 (V_4^2 - V_3^2) +
              R^4 -4 \hbar r R^2 V_1 $\cr} \; ,
$$
$$
(\vert z_1 \vert^2 -R^2)(\vert z_2 \vert^2 -R^2)
=\cases{ &{\-}$(2r \hbar A_3 -R^2)^2 - 4 \hbar^2 r^2 A_4^2 $\cr
        & ${\-}$ \cr
        &{\-}$ (2r \hbar V_4 -R^2)^2 - 4 \hbar^2 r^2 V_3^2 $  \cr}
$$
involving explicitly the algebra generators,
clearly exhibit the absence of any ordering problem.
The expressions constituting the denominators of ${\cal A}^{-1}$
are in fact linear combinations of powers of the generators
of the algebra involved in the two-particle quantization scheme.
The explicit form of the operator ${\cal A}^{-1}$ is easily
obtained by resorting to the Laplace integral representation
of the function $f(x) = 1/x$ (see Ref. \cite{PE1}). 
%
%
\vskip 0.5 truecm
\centerline{\bf APPENDIX B}
\vskip 0.5 truecm
The action of $R_{\pm}$ on the canonical variables $x_j$, $p_j$
can be derived from the formulas
$$
U_{\psi} \,
\left( \matrix{ x_1& \cr x_2& \cr p_1& \cr p_2& \cr} \right) \,
U_{\psi}^{+} \,=
\cases{ 
&${x_1 \, {\rm cos} \psi + x_2 \,{\rm e}^{\mu}{\rm sin} \psi }$ \cr
&${\-}$ \cr 
&${x_2 \, {\rm cos} \psi - x_1 \,{\rm e}^{-\mu} {\rm sin} \psi }$ \cr 
&${\-}$ \cr 
&${p_1 \,{\rm cos} \psi + p_2 \,{\rm e}^{-\mu} {\rm sin} \psi }$ \cr
&${\-}$ \cr 
&${p_2 \, {\rm cos} \psi -p_1 \,{\rm e}^{\mu} {\rm sin} \psi }$ \cr}
$$
where $U_{\psi} ={\rm exp}(-i 2 \psi V_2)$.  To this aim, it is
important to consider the two transformations
$$
{\cal U}_j(\mu) x_j {\cal U}_j^{\dagger}(\mu) = 
{\rm e}^{\mu} x_j \quad , \quad
{\cal U}_j(\mu) p_j {\cal U}_j^{\dagger}(\mu) = 
{\rm e}^{-\mu} p_j \, ,
$$
where ${\cal U}_j (\mu) = {\rm exp}
[i\mu (x_j p_j +p_j x_j)/(2 \hbar) ]$,
involving the two decompositions
$$
U_{\psi} ={\cal U}_2(\mu) e^{-i2 \psi L_3} {\cal U}_2^{\dagger}(\mu)
\, , \;
U_{\psi}={\cal U}_1^{\dagger}(\mu) e^{-i2 \psi L_3} {\cal U}_1(\mu)
\, ,
$$
with $L_3 = (x_1 p_2 - x_2 p_1)/ \hbar$. 
Such decompositions and the fact that 
${\cal U}_j(\mu)\,\Psi_n (x_j; r_j) =
\Psi_n (x_j ;r_j {\rm e}^{-2\mu})$
(see, f.i., Ref. \cite{PE3}),  allows one to obtain the
explicit form of eigenfunctions (\ref{EF1}) and (\ref{EF2}).
When assuming ${\rm e}^{\mu} = {\sqrt {r_1 /r_2}}$ one finds
$$
\Psi_{nq}^{+} (x_1, x_2) = \Psi_q (\alpha_2 X ; r_2) \,\Psi_n
(x/\alpha_2 ; r_1) \; ,
$$
$$
\Psi_{nq}^{-} (x_1, x_2) = \Psi_q (\alpha_1 X ; r_1) \,\Psi_n
(-x/\alpha_1 ; r_2) \; ,
$$
respectively, where $\alpha_j = \sqrt{C/\rho k_j}$, and
coordinates (\ref{CO1}), (\ref{CO2}) have been used.
%
%
%
%
\vskip 0.5 truecm
\centerline{\bf APPENDIX C}
\vskip 0.5 truecm
\noindent
Replacing $r_2$ with $-r_2$ in $V_3$, $V_4$ and $p_2$ with
$-p_2$ in $V_1$, $V_2$ [see formula (\ref{AG3})-(\ref{AG2})
and (\ref{AG4})] provides the su(1,1) operators $A_4$, $A_3$
and $A_1$, $A_2$, respectively. The final form
$$
A_3= {\frac{r}{8\hbar}} \left [\xi^2 + \eta^2 + 
{\frac{4}{r^2}} ({\cal X}^2 + {\cal Y}^2) \right ] \;,
$$
$$
A_1= {\frac{r}{8\hbar}} 
\left [{\frac{4}{r^2}} ({\cal X}^2 + {\cal Y}^2)
-(\xi^2 + \eta^2 ) \right ] \;,
$$
$$
A_2= -{\frac{1}{2\hbar}}(\xi {\cal X} + \eta {\cal Y}) \; ,\;
A_4={\frac{1}{2\hbar}}(\xi {\cal Y} - \eta {\cal X}) \, ,
$$
is achieved when the special coordinates (\ref{VAV}) of the case 
$k_2=k_1$ are employed.
%
%

%

\begin{references}
%
\bibitem{BO} A. L. Fetter,  Phys. Rev. {\bf 162}, 143 (1967).
%
\bibitem{RARE}
M. Rasetti and T. Regge, Physica A{\bf 80}, 217, (1975);
G. A. Goldin, R. Menikoff, and D. H. Sharp, 
Phys. Rev. Lett. {\bf 67}, 3499 (1991);
V.Penna and M. Spera, J. Math. Phys. {\bf 33}, 901 (1992);
Y. Wu, ibid {\bf 34}, 2342 (1993).
%
\bibitem{AHNS} V. Ambegaokar, B. I. Halperin, D. R. Nelson,
and E. D. Siggia, Phys. Rev. B {\bf 21}, 1806 (1980).
%
\bibitem{CHM} R. Y. Chiao, A. Hansen, and A. Moulthrop,
Phys. Rev. Lett. {\bf 55}, 2887 (1985).
%
\bibitem{HW} F. D. M. Haldane and Y.-S. Wu, 
Phys. Rev. Lett. {\bf 55}, 1431 (1985).
%
\bibitem{LEMY} J. Leinaas and J. Myrheim, 
Phys. Rev. B {\bf 37}, 9286 (1988).
%
\bibitem{GMS2} G. A. Goldin, R. Menikoff, and D. H. Sharp, 
Phys. Rev. Lett. {\bf 58}, 174 (1987). 
%
\bibitem{ES} U. Eckern and A. Schmid,
Phys. Rev. B {\bf 39}, 6441 (1989).
%
\bibitem{PE1} V. Penna,  Physica A {\bf 152}, 400 (1988).
%
\bibitem{PE2} V. Penna,  Phys. Lett. {\bf 125A}, 385 (1987).
%
\bibitem{LE} J. M. Leinaas, Ann. Phys. (N.Y.) {\bf 198}, 24 (1990).
%
\bibitem {TG} A. M. Thompson and J. M. F. Gunn,
Physica C {\bf 235}, 2953 (1994).
%
\bibitem{TITI}
{\it Superfluidity and Superconductivity}, edited 
by D. R. Tilley and J. Tilley (Hilger, London, 1986), 2nd ed.; 
%
\bibitem{KT1} J. M. Kosterlitz and D. J. Thouless,
J. Phys. C  {\bf 6}, 1181 (1973).
%
\bibitem{MIN} P. Minnhagen, Rev. Mod.. Phys. {\bf 59}, 1001 (1987)
%
\bibitem{NAT} Q. Niu, P. Ao, and D. J. Thouless,
Phys. Rev. Lett. {\bf 72}, 1706 (1994).
%
\bibitem{SUSU}
{\it Proceedings of the 20th International Conference on
Low Temperature Physics}, edited by R. J. Donnelly,
[Physica B {\bf 194-196}, (1994)].
%
\bibitem{MINA} T. Mingoguchi and Y. Nagaoka, 
Prog. Theor. Phys. {\bf 80}, 397 (1988).
%
\bibitem{TAN} D. J. Thouless, P. Ao, and Q. Niu,
Physica A {\bf 200}, 42 (1993).
%
\bibitem{IEJU} R. Iengo and G. Jug,
Phys. Rev. B {\bf 54}, 13207 (1996).
%
\bibitem{DSSM} J. C. Davis, J. Steinhauer, K. Schwab,
Yu. M. Mukharsky, A. Amar, Y. Sasaki, and R. E. Packard, 
Phys. Rev. Lett. {\bf 69}, 323 (1992); 
G. G. Ihas, O. Avenel, R. Aarts,
and R. Salmelin, ibid. {\bf 69}, 327 (1992)
%
\bibitem{JJA} Vortex dynamics and Quantum effects
in Josephson junction arrays are reviewed in the Proceedings of the 
ICTP Workshop on {\it Josephson Junction Arrays},
edited by H. A. Cerdeira 
and S. R. Shenoy [Physica B {\bf 222}, 336-370 (1996)]. 
%
\bibitem{DON}
R. J. Donnelly, {\it Quantized Vortices in He II},
(Cambridge University Press, Cambridge, 1991). 
%
\bibitem{LEGU} H. H. Lee and J. M. F. Gunn,
Phys. Rev. B {\bf 46}, 8336 (1992).
%
\bibitem{LUSI} R. Lupini and S. Siboni,
Nuovo Cimento A {\bf 106}, 957 (1991).
%
\bibitem{ZFG} W. M. Zhang, D. H. Feng, and R. Gilmore,
Rev. Mod. Phys. {\bf 62}, 761 (1990).
%
\bibitem{ONS} L. Onsager, Nuovo Cimento Suppl. {\bf 6}, 249 (1949).
%
\bibitem{FO} R. P. Feynman,
in {\it Progress in Low-Temperature Physics}, edited by
C. J. Gorter  (North-Holland,  Amsterdam, 1955), Vol. 1, p. 17. 
%
\bibitem{SAF} P. G. Saffman, {Vortex Dynamics}
(Cambridge Univ. Press,  Cambridge, 1992).
%
\bibitem{ARE}
H. Aref, Ann. Rev. Fluid Mech. {\bf 15}, 345 (1983)
%
\bibitem{BPS} V. Barone, V. Penna and P. Sodano,
Ann. Phys. (N.Y.) {\bf 225}, 212 (1993).
%
\bibitem{PE3} V. Penna, Ann. Phys. (N.Y.) {\bf 245}, 389 (1995).
%
\bibitem{ABST} {\it Handbook of Mathematical Functions},
edited by M. Abramowitz and I. A. Stegun (Dover, New York, 1972).
%
\bibitem{MAG} R. E. Prange and S. M. Girvin, {\it The Quantum Hall
Effect} (Springer-Verlag, New York, 1987);
A. Hansen, E. H. Haunge J. Hove and F. A. Maas,
In {\it Annual Review of Computation Physics}, Vol. {\bf 5},
edited by D. Stauffer (World Scientific, Singapore, 1997).
%
\bibitem{BER} A. Shapere and F. Wilczek,
{\it Geometric Phase in Physics} (World Scientific, Singapore, 1992).
%
\bibitem{MOFE} P. M. Morse and M. Feshbach,
{\it Methods of Theoretical Physics}, Vol. 2
(McGraw Hill, New York, 1970).
%
\bibitem{LINA} G. Lindblad, and B. Nagel, Ann. Inst. Henri Poincar\'e,
{\bf 13}, 27 (1970).
%
\bibitem{GRA} I. S. Gradshteyn and I.M. Ryzhik, 
{\it Table of Integral, Series, and Products} (Academic Press,
New York, 1980).
%
\bibitem{PERA} V. Penna, and M. G. Rasetti,
Phys. Rev. B {\bf 50}, 11783 (1994).
%
\bibitem{MOPE} A. Montorsi, and V. Penna,
Phys. Rev. B {\bf 55}, 8226 (1997).
%
\end{references}
\end{document}